\documentclass[preprint,showpacs,preprintnumbers,amsmath,amssymb,prl]{revtex4}
\usepackage{amsmath,amssymb,amsfonts}
\usepackage{mathrsfs}
\usepackage{graphicx}
\usepackage{dcolumn}
\usepackage{bm}

\begin{document}

\title{Evidence of light guiding in ion-implanted diamond}
\author{S. Lagomarsino}
\email{lagomarsino@fi.infn.it}
\affiliation {Department of Energetics University of Florence, \\ Via S.Marta 3, 50136 Italy}
\altaffiliation{INFN sezione di Firenze, Via Bruno Rossi, 50019 Sesto Fiorentino (FI) Italy}
\author{F. Bosia}
\affiliation {Experimental Physics Department and "Nanostructured Interfaces and Surfaces" Centre of Excellence, University of Torino, via P. Giuria 1, 10125 Torino, Italy}
\altaffiliation{INFN Sezione di Torino, via P. Giuria 1, 10125 Torino, Italy}
\author{M. Vannoni}
\affiliation{CNR, Istituto Nazionale di Ottica Applicata (INOA), Largo E. Fermi 6, 50125 Arcetri, Firenze Italy}
\author{S. Calusi}
\author{L. Giuntini}
\author{M. Massi}
\affiliation{Physics Department and INFN Sezione di Firenze, University of Firenze (Italy), via Sansone 1, 50019 Sesto Fiorentino, Firenze, Italy}
\author{P. Olivero}
\affiliation {Experimental Physics Department and "Nanostructured Interfaces and Surfaces" Centre of Excellence, University of Torino, via P. Giuria 1, 10125 Torino, Italy}
\altaffiliation{INFN Sezione di Torino, via P. Giuria 1, 10125 Torino, Italy}

\begin{abstract}
The refractive index of a monocrystalline diamond has been significantly increased along buried micro-channels by means of  proton implantations using a scanning focused micro-beam, leading to the direct fabrication of 50 $\mu$m deep waveguides below the diamond surface.
A phase-shift micro-interferometric method has been applied to the observation of the fabricated structures, obtaining evidence of light propagation in guided modes inside the modified regions 
\end{abstract}

\maketitle
Diamond color defects have been indicated as promising features to be used in next quantum computing devices; recently, nitrogen-vacancy defects, nickel, silicon and chromium based complexes have been used for single photon sources and quantum bit storage elements at room temperature \cite{Beveratos2002, Kok2006, Prawer2008, Aharonovich2009, Chunlang2006, Aharonovich2010}. A reliable fabrication method of monolithic photonic devices in diamond is therefore a required step to further  develop the research field on the path to fabricate future monolithic quantum devices \cite{Gu2004, Hiscocks2008, Hiscocks2008a, Wang2007, Fairchild2008}.

All of the methods employed so far in the micro-fabrication of diamond for photonic applications exploit the refractive index contrast between diamond and air, and rely on material ablation to create structures such as photonic cavities and waveguides. 
In comparison, direct ion micro-beam writing\cite{Bettiol2005}, a technique extensively employed in a range of materials including polymers\cite{Sum2003} and glasses\cite{Bettiol2006}, could offer unique opportunities for a more rapid prototyping of micro-photonic devices, exploiting also a refraction index contrast much lower than that between diamond and air, which can be produced by ion irradiation damage. Moreover, since the same methodology (with much higher irradiation fluences) can be employed in diamond for fabrication of structures such as microfluidic channels, direct beam writing could open the way 
to biosensing applications through the production of ``lab-on cells'' where the cells can be chemically and/or optically stimulated and their optical responses collected by means of an integrated optics.

In this letter we report on a method to create local modifications of the optical properties of diamond, as very small refractive index variations buried in the diamond bulk material, using a MeV proton micro beam.
We exploited our previous extensive work on the optical modifications induced in diamond by 3 MeV proton irradiation\cite{Olivero2009},  to obtain controlled increments of the refractive index in rectilinear, 500 $\mu$m-long, 12 $\mu$m-wide structures below the diamond surface, down to a depth of about 50 $\mu$m. 

In other contexts, microscopic images are exploited to indirectly calculate the variations in refractive index induced by ion irradiation\cite{Bettiol2006, Bibra1997}. We use a quite different approach: we are able to design a refractive index profile in two or three dimensions and to realize it by micro-beam writing, then we can measure directly the modes amplitude profile with a phase-shift micro-interferometric technique and compare it with the modes amplitudes previously calculated on the basis of the designed refractive index distribution. 

\begin{figure}
\includegraphics [width=0.8\textwidth]{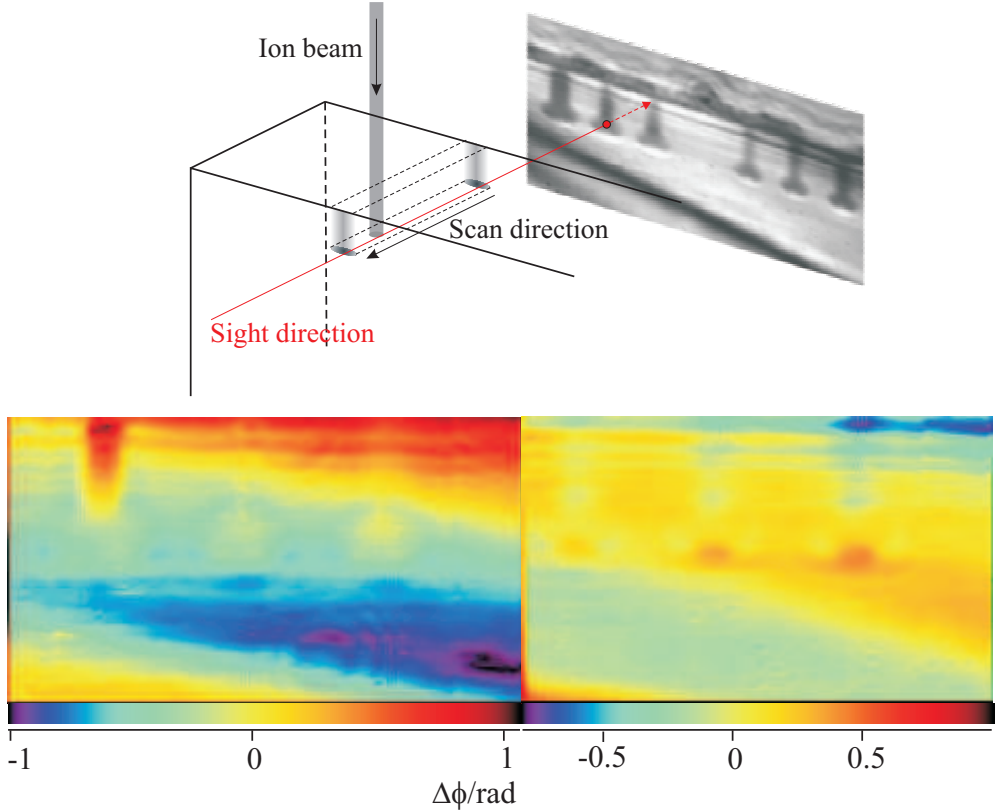}
\caption{\label{fig: 1} (top) Scheme of the implantation geometry and interference pattern produced by the implantation at (from left to right) $2\times 10^{16}$cm$^{-2}$ fluence (one implantation), $1\times 10^{16}$cm$^{-2}$ (two implantations), $5\times 10^{15}$cm$^{-2}$ (the last three implantations). (down) Phase maps corresponding to the same implantations.}
\end{figure}

\begin{figure}
\includegraphics [width=0.8\textwidth]{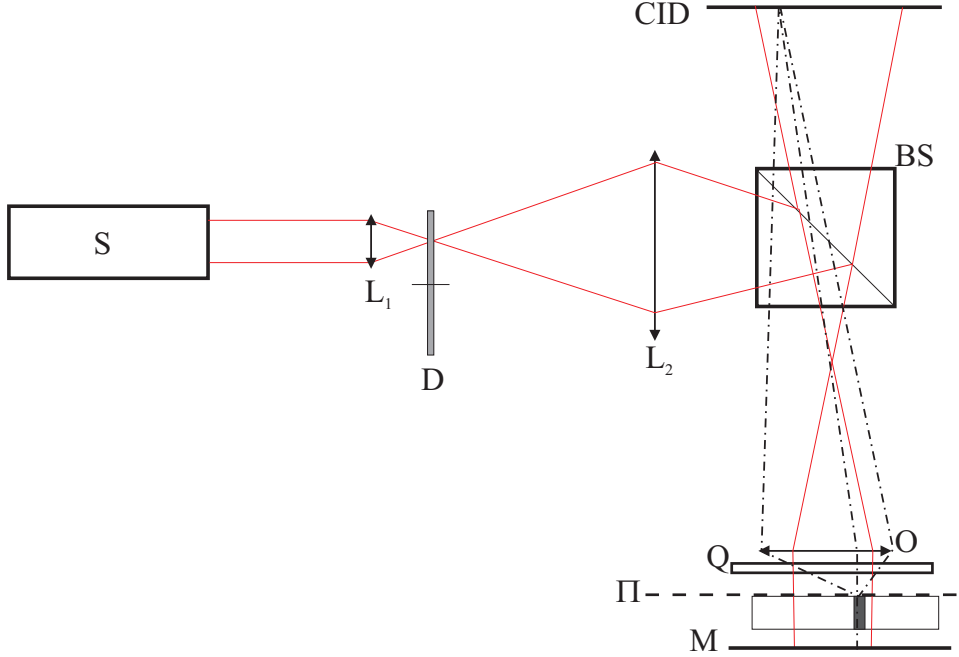}
\caption{\label{fig: 2} Scheme of the principle of measurement of the micro-interferometer (see text for the explanation)}
\end{figure}

To perform this study, three adjacent faces of a $3.0\times 3.0\times 0.5$ mm$^3$ sample of type IIa single-crystal chemical-vapor deposited diamond, cut along the $\langle 100\rangle$  axes, were optically polished down to a roughness of some nanometers, and then implanted at the external scanning microbeam facility of the LABEC laboratory in Florence\cite{Giuntini2007} . The beam was focused on the small polished side of the sample to an approximately Gaussian spot of 12 $\mu$m width, and scanned along a rectilinear path perpendicular to the two polished faces of the largest size at implantation fluences of $2\times 10^{16}$, $1\times 10^{16}$, $5\times 10^{15}$cm$^{-2}$ in the central region, with an estimated uncertainty not exceeding 5\% (see the top of Fig.\ref{fig: 1} for a sketch of the implantation set-up).
The as-prepared structures were then observed with a commercial laser interferometric microscope (Maxim 3D Zygo, Connecticut, USA, www.zygo.com) along the longitudinal direction of the guide. The setup used has been proven useful to characterize refractive index variations by observation of the phase shift induced by thin planar structures, in order to perform optical path difference (OPD) measurements. Here we claim it to be useful for a different kind of measurement, since the phase information of the field emerging from the narrow, long structures implanted in diamond can give indication about the amplitude of the guided modes. For this reason, it is useful to resume the principle of the measurement in relation to our purposes (see also Fig.\ref{fig: 2}).

An He-Ne laser (S) beam  is focused on a diffusion disk (D) with the lens L$_1$ to obtain a spatially incoherent, s-polarized source that is focused again by the lens L$_2$ on a polarizing beam splitter BS. The light illuminates a microscope objective O, then the collimated beam invest the quarter-waveplate Q, whose last surface reflects back about 10\% of the incident beam intensity, to act as a reference wavefront. The microscope objective focuses the light to cross through the sample, reflecting from an high quality mirror M back through the sample and the quarter waveplate, becoming a p-polarized beam that is added to the reference beam on a charge injected device camera sensor (CID). The reference and measuring wavefronts produce an interference pattern that can be recorded by the electronic system. A piezo-electric micro actuator is able to displace the reference surface with $\lambda_{\text{He-Ne}}/8$ steps taking at least 5 consecutive intensity measurements on the CID plane, and a phase-shift algorithm is used to reconstruct variation in the phase of the measured electric field on the plane $\Pi$ with an accuracy of about 2$\pi$/1000 rad.

Figure \ref{fig: 1} (upper image) reports an image, in perspective, of the interference pattern produced on the CCD plane by the $2\times 10^{16}$cm$^{-2}$ implantation, by two $1\times 10^{16}$cm$^{-2}$ and by three $5\times 10^{15}$cm$^{-2}$implantations (from the left to the right), Fig.\ref{fig: 1} (lower image) represents the same areas in the false-color phase maps reconstructed by the computerized system of the instrument. We show now that these phase maps can be interpreted as a direct measurement of the amplitudes of the modes propagating along the guide.

Let the electric field on the plane $\Pi$ (in the polarization direction) be described by the real part of the function
\begin{equation}
E=\mathscr{E}\left( x,y\right)e^{i\left[\omega t+\phi\left( x,y\right) \right]}.
\end{equation}
If the structures under consideration have a cross-sectional dimension comparable to that of the wavelength of the radiation,
the radiation emerging from the diamond will be given by a principal plane-wave part plus a perturbation produced by the structures themselves. Consequently, the field on the plane $\Pi$ will be given by the sum of three contributions: a principal part given by the back radiation reflected by the mirror M
\begin{equation}
E_0=\mathscr{E}_0 e^{i\left(\omega t+\phi_0 \right)},
\end{equation}
a secondary field deriving from the reflections on the surfaces of the sample, which is
\begin{equation}
E_R =\mathscr{E}_R\left( x,y\right)e^{i\left[\omega t+\phi_R\left( x,y\right) \right]},
\end{equation}
and a perturbation given by the contribution to the field of the structures under consideration. If the field can be considered as guided by the structures, this contribution can be simply written as
\begin{equation}
E_G =f\left( x,y\right)e^{i\left(\omega t+\phi_G \right)},
\end{equation}
were the function $f\left( x,y\right)$ is the amplitude map of the mode.

If $\mathscr{E}_R$ and $f$ are both small compared with $\mathscr{E}_0$, the phase difference $\Delta\phi =\phi\left( x,y\right)-\phi_0$, measured by the instrument, is given, at the lowest order, by: 
\begin{eqnarray}
\Delta\phi\left( x,y\right)= &&\frac{\mathscr{E}_R\left( x,y\right)}{\mathscr{E}_0} \sin\left( \phi_R\left( x,y\right)-\phi_0\right)
\nonumber
\\
&&+\frac{f\left( x,y\right)}{\mathscr{E}_0} \sin\left( \phi_G-\phi_0\right).
\end{eqnarray}
Consequently, once the contribution of the reflections has been fitted and subtracted, the map of $\Delta\phi$ is simply proportional to the amplitude map of one of the modes which can propagate in the structure, or to a linear combination of several modes simultaneously propagating in the waveguide, each with its appropriate phase value $\phi_G$.

\begin{figure}
\includegraphics [width=0.8\textwidth]{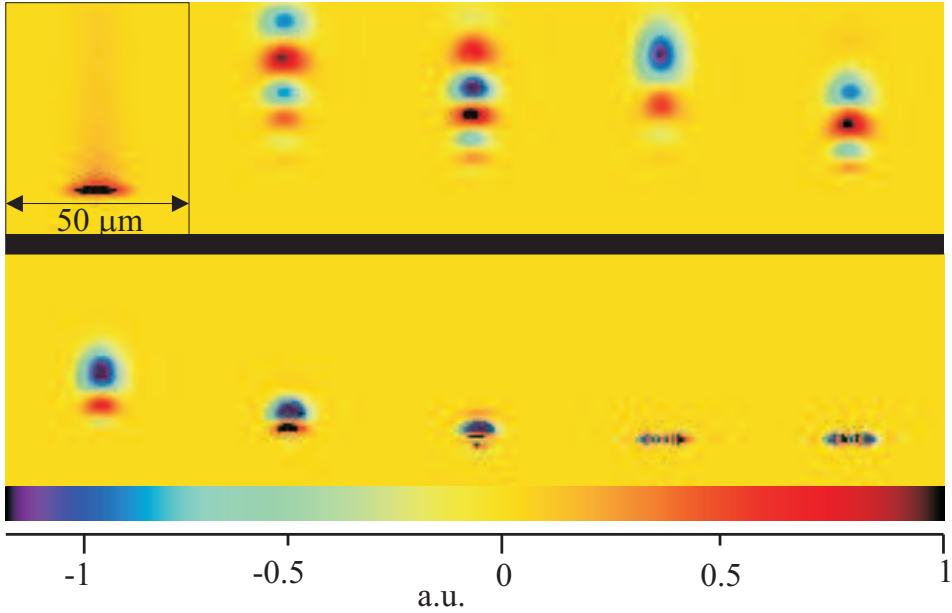}
\caption{\label{fig: 3} (top left) Map of the damage produced by the ion beam in term of vacancy density. (rest of the figure) In the same spatial scale, amplitude maps of 9 simulated modes propagating in the waveguide irradiated at $2\times 10^{16}$cm$^{-2}$ fluence.}
\end{figure}

In order to compare the experimental phase profiles with a superposition of modes propagating in the wave\- guides, a 2-dimensional finite element model (FEM) of the irradiated regions was employed, taking into account the local modifications in the refractive index induced by proton damage, quantified in terms of the induced vacancy density  and calculated by means of a Monte Carlo simulation (SRIM®) as reported in ref.\cite{Olivero2009}. Once given the vacancy density at every cell of the simulation grid, the local variation of refractive index at the He-Ne wavelength of 632.8 nm is calculated on the basis of the simple linear relation, given by ref.\cite{Olivero2009} for the real part, while the imaginary part follows from data to be published in a forthcoming work. The result is
\begin{equation}
\Delta n= \left( 4.20 + 2.88 i\right)\times 10^{-23}\frac{dN_{vacancies}}{dV/\text{cm}^3}.
\label{ref_index}
\end{equation}
FEM simulations were carried out using the RF module of COMSOL Multiphysics \- (http://www.comsol.com). Specimen geometry was reproduced and meshed using quadrilateral elements, using Eq.\ref{ref_index} for the refractive index variation in the adjacent waveguides. The value used for the refractive index of unimplanted diamond was $n_0=2.41$. Simulations yield the effective mode index (i.e. the propagation constant) of the confined modes in the waveguides and the corresponding electric and magnetic field amplitude distributions. These data can be compared to the experimentally measured maps. 
In figure \ref{fig: 3} a map of the vacancy density, as calculated by SRIM, and an example of 9 different numerically calculated modes propagating in the structures irradiated at the higher fluence are shown. At lower fluencies, progressively lower refractive index modifications imply a lower number of propagating modes, whose amplitude maps present fewer and wider lobes.

\begin{figure}
\includegraphics [width=0.8\textwidth]{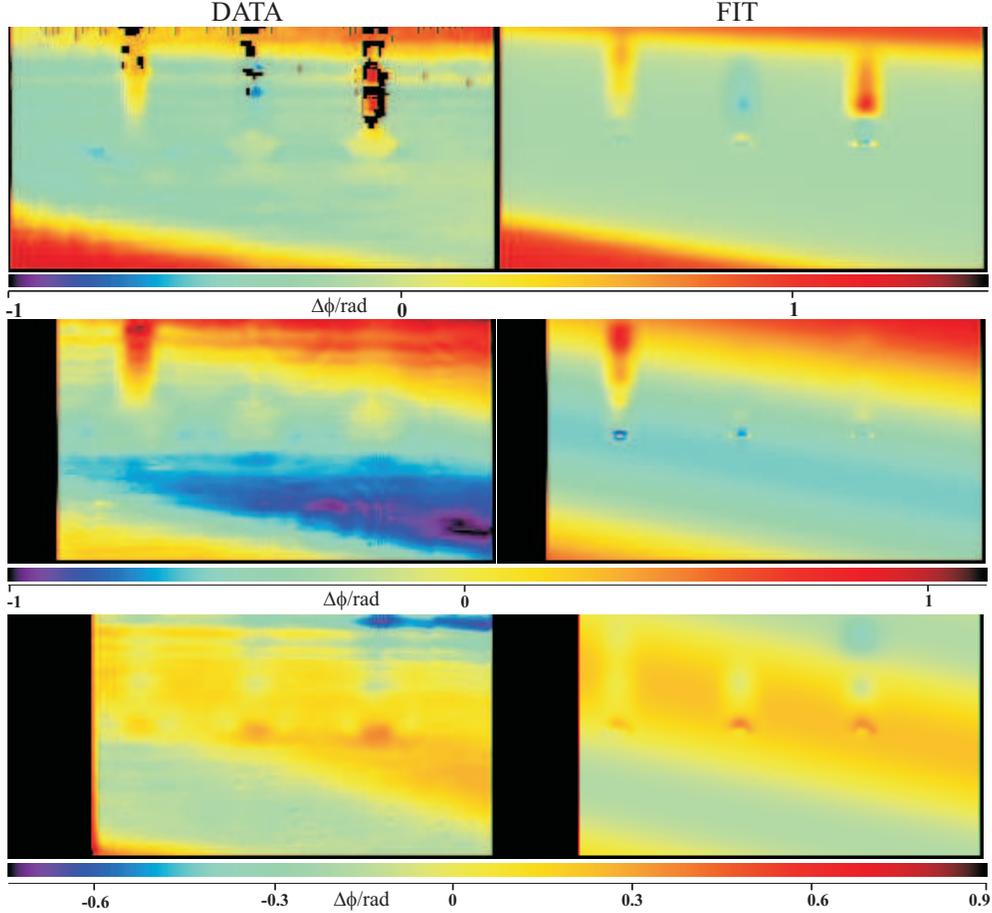}
\caption{\label{fig: 4} Comparison of the measured phase shift maps (left) with the fit (right) obtained by linear superposition of modes amplitudes and a background taking into account multiple reflections effects. (top and middle) Images related to three adjacent guides obtained with irradiation at $2\times 10^{16}$cm$^{-2}$ (the one at the left) and at $1\times 10^{16}$cm$^{-2}$ (the others). (bottom) Images related to three guides irradiated at $5\times 10^{15}$cm$^{-2}$ of fluence.}
\end{figure}

The experimentally obtained phase maps were compared with a superposition of the calculated amplitude maps, by fitting them with a linear combination of the propagating modes. Since the relative amplitudes of the modes excited in the waveguides depend in a sensitive way from the illumination conditions, different positions of the sample on the focal plane may imply different weights to be assigned at each particular mode. 
In figure \ref{fig: 4} two different images (top and middle images) of the same implantations at fluences of 2 and $1\times 10^{16}$cm$^{-2}$ are shown along with the best fit (on the right) obtained with 30 different propagation modes (10 with the maximum on the first structure, 10 on the second and 10 on the third) and a combination of sinusoids taking into account the background given by the reflections on the two diamond plane surfaces. It is evident that the same set of propagation modes, though with different weights, fits the two images. The weight of each mode is not of much significance, since it is proportional to the sinus of $\phi_G-\phi_0 $, an angle which is supposed to vary randomly from a mode to another. In the same figure (bottom), an image of the $5\times 10^{15}$cm$^{-2}$ fluence waveguide is also shown, together with its 21-modes fit.

From the inspection of these images we conclude that the adherence of the fit to the experimental bi-dimensional profiles is very good in the cap layer between 0 and about 45 $\mu$m in depth, where the relative damage is small, while at end-of-range the structures seem to be more diffuse, probably due to the distortion induced by diffraction on the highly opaque regions, which lies under the plane $\Pi$ (see Fig.\ref{fig: 2}) in correspondence with the considered structures.

In conclusion, there is sufficient evidence that the ion-beam writing method proposed in this letter gives the possibility to fabricate light-guiding structures in bulk diamond. Moreover, the micro-interferometric measuring method briefly exposed here provides means for a highly detailed study of the mode patterns propagating in the waveguides.

%


\end{document}